\documentclass[fleqn,usenatbib]{mnras}
\usepackage[T1]{fontenc}
\usepackage{ae,aecompl}
\usepackage{graphicx}	
\usepackage{amsmath}	
\usepackage{amssymb}	

\title{A Correlation in the Waiting-time Distributions of Solar Flares} 

\author[Hudson]{
\begin{tabular}{lll}
Hugh S. Hudson$^{1,2}$
\end{tabular}
\bigskip
\\
$^{1}$School of Physics and Astronomy, U. of Glasgow, UK\\
$^{2}$Space Sciences Laboratory, U.C. Berkeley, CA USA
}

\begin{document}
\maketitle

\begin{abstract}
In isolated solar active regions, we find that the waiting times between flares correlate with flare magnitudes as determined by the GOES soft X-ray fluxes.
A ``build-up and release'' scenario (BUR) for magnetic energy storage in the solar corona suggests the existence of
such a relationship, relating the slowly varying subphotospheric energy sources to the sudden coronal energy releases of flares and CMEs.
Substantial amounts of research effort had not previously found any obvious observational evidence for such a BUR process.
This has posed a puzzle since coronal magnetic energy storage represents the consensus view of the basic flare mechanism.
We have revisited the GOES soft X-ray flare statistics for any evidence of correlations, using two isolated active regions, 
and have found significant evidence for a ``saturation'' correlation.
Rather than a ``reset'' form of this relaxation, in which the time \textit{before} a flare correlates with its magnitude, the  ``saturation'' relationship results in the time \textit{after} the flare showing the correlation.
The observed correlation competes with the effect of reduced GOES sensitivity, in which weaker events can be under-reported systematically.
This complicates the observed correlation, and we discuss several approaches to remedy this.
\end{abstract}


\section{Introduction}

Solar flares appear to result from the build-up of ``free'' magnetic energy in the corona, which can grow slowly as the result of non-radiative energy fluxes (such as wave energy or Poynting flux) injected into the solar atmosphere.
These excess energies appear as non-potential field structures, describable in terms of electrical currents that the corona to the solar interior. 
A flare (or a CME) extracts energy suddenly from this stressed field, which then must relax to a lower-energy state.
This well-understood scenario has broad acceptance, if only because alternative mechanisms for energy storage seem so implausible observationally \citep[e.g.,][]{2007ASPC..368..365H}.
This basic picture spawned a major effort in the 1980s, the ``Flare Build-up Study'' \citep{1987SoPh..114..389G}; its first objective was to identify signatures of the sequence of slow buildup and rapid flare energy release.
This did not happen: ``...no consistent relationship was found.. ,'' providing a motivation for the present work.
Nowadays we have substantially improved observational material for characterization and for follow-up.
Some of the non-potential ``free energy'' appears to remain in the corona even after a major flare \citep{1994ApJ...424..436W}. 
We note that more numerous weak flares or even steady heating can also derive from the reservoir of coronal free energy.

A system exhibiting slow buildup and rapid release (BUR) constitutes a ``relaxation oscillator.''
The driven ``oscillation'' need not be periodic, as was noticed at the very outset \citep{1927Natur.120..363V}, and can exhibit chaotic behavior (``irregular noise''); see \cite{2012Chaos..22b3120G} for the full history.
Many natural systems exhibiting this kind of periodic behavior exist, as do engineering applications; in astrophysics the ``Rapid Burster'' X-ray source gave an early example \citep{1976ApJ...207L..95L}.
In gardens with water features one can find often a mechanical ``dipper'' with a trickle of flowing water; this operates on the same BUR principle.
The dipper form of a BUR process corresponds to a ``reset'' of the free energy as the bucket empties and the slower refilling starts again; this predicts a correlation between the time to build up the event energy and the magnitude to reset it to zero.
In the case of a steady input, such as the trickle of water in the dipper, the regular resets create a regular oscillation, but other variants of this toy model do exist.

\cite{1978ApJ...222.1104R} gave an early theoretical description of such a BUR process for solar flares.
This inspired considerable literature on flare occurrence distributions, none of which appears to have succeeded in an experimental confirmation of this kind of relationship.
Flare occurrence generally follows a power-law distribution of magnitudes, much as earthquakes and many other natural systems do; this suggests a self-organized critical phenomenon \citep[see, e.g.,][for a recent review]{2016SSRv..198...47A}.
The waiting times between successive flares appear to follow a ``piecewise Poisson'' random pattern \citep{2000ApJ...536L.109W}.

To establish the BUR process we essentially need to compare the free energy content of the corona, and we need an estimation of the energy release in each event, in terms of observable quantities.
Both tasks are difficult, but we can finesse the measurement of stored energy by assuming that it changes only slowly, on time scales greater than those of the flare release.
Many proxies for flare energy exist, and we are helped by the tendency of many of the extensive parameters to scale together (one manifestation of the ``Big Flare Syndrome''; Kahler, 1982).
\nocite{1982JGR....87.3439K}
In previous efforts to establish a BUR correlation, \cite{1994AIPC..294..183B} used hard X-ray observations from the BATSE experiment \citep{1992NASCP3137...26F},  \cite{1998ASSL..229..237H} used the GOES soft X-ray observations, and both \cite{1998A&A...334..299C} and \cite{2000SoPh..191..381W}  used data from the WATCH monitor of hard X-rays, as described in an early version by \cite{1981Ap&SS..75..145L}.
The relationship of any of these proxies to the total flare energy certainly contains some variance, and none represents a large fraction of the total.
Proxies for the total luminous energy of a flare also miss the energy of any associated coronal mass ejection (CME), a major factor for a minority of flares.

None of the earlier studies have reported a significant interval-size relationship, which would observationally support a BUR process if detected.
This paper again uses the standard GOES soft X-ray database (Section~\ref{sec:data}) but specifically seeks an alternative form of the BUR correlation, as described in Section~\ref{sec:relationship}.
The observations clearly show an effect consistent with the BUR idea, but a systematic feature of the GOES time series complicates the picture (Section~\ref{sec:interp}). 
We suggest further specific searches to clarify this situation.

\section{Database}\label{sec:data}

We again use the GOES soft X-ray proxy for flare energy, plus the metadata regarding the identification of the flaring active region.
This is the standard material available from SolarSoft \citep{1998SoPh..182..497F}.
The GOES event classification (e.g., ``X1.2'' standing for $1.2 \times 10^{-4}$~W/m$^2$) represents irradiance (flux), rather than fluence, whereas a more appropriate proxy for total flare energy might have units of energy rather than power.
\cite{1995PASJ...47..251S} suggested a soft X-ray energy fraction of order 1\%, for example, but this fraction has an unknown variance and also must have, based upon models, some systematic bias across the scale of flare magnitudes.
However the peak GOES irradiance correlates with the flare soft X-ray fluence, and so it also can serve as a proxy.
Even the GOES fluence itself has a systematic bias resulting from its temperature weighting, since the actual data
consist of broad-band samples centered at photon energies well above flare $kT$ values determined spectroscopically.

\begin{figure}
\centering
\includegraphics[width=\columnwidth]{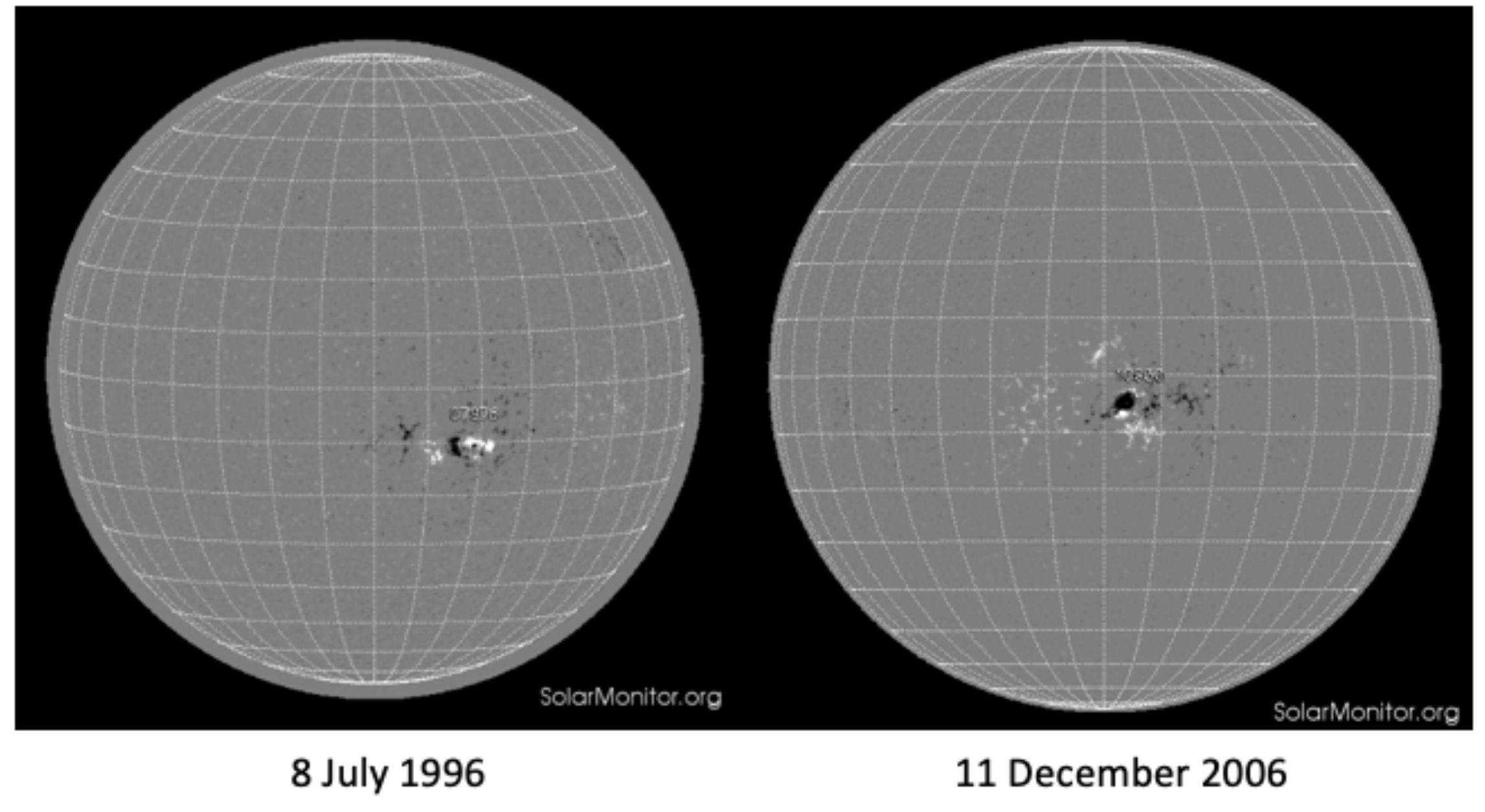}
     \caption{\textit{File magnetograms for the two  regions under study, courtesy SolarMonitor.
        }
} 
\label{fig:magnetograms}
\end{figure}

This study covers two active regions, NOAA~07978 (July, 1996) and NOAA~10930 (December, 2006).
Each of them exclusively made all of the GOES flares for the intervals studied, and as can be seen from the file magnetograms in Figure~\ref{fig:magnetograms}, each was truly isolated (for noting this about NOAA~10930, we thank M.~Georgoulis, personal communication 2019).
They both produced remarkably powerful flares near the very end of their sunspot cycles, and in each case the flares
had singular properties that their isolation may have helped to make detectable.
SOL1996-07-09 (X2.2) spawned the first observed ``sunquake'' \citep{1998Natur.393..317K};
SOL2006-12-13 (X3.4) followed the event producing the first detectable MeV-energy flux of neutral atoms \citep{2009ApJ...693L..11M}.

\begin{figure*}
\centering
\includegraphics[width=1.5\columnwidth]{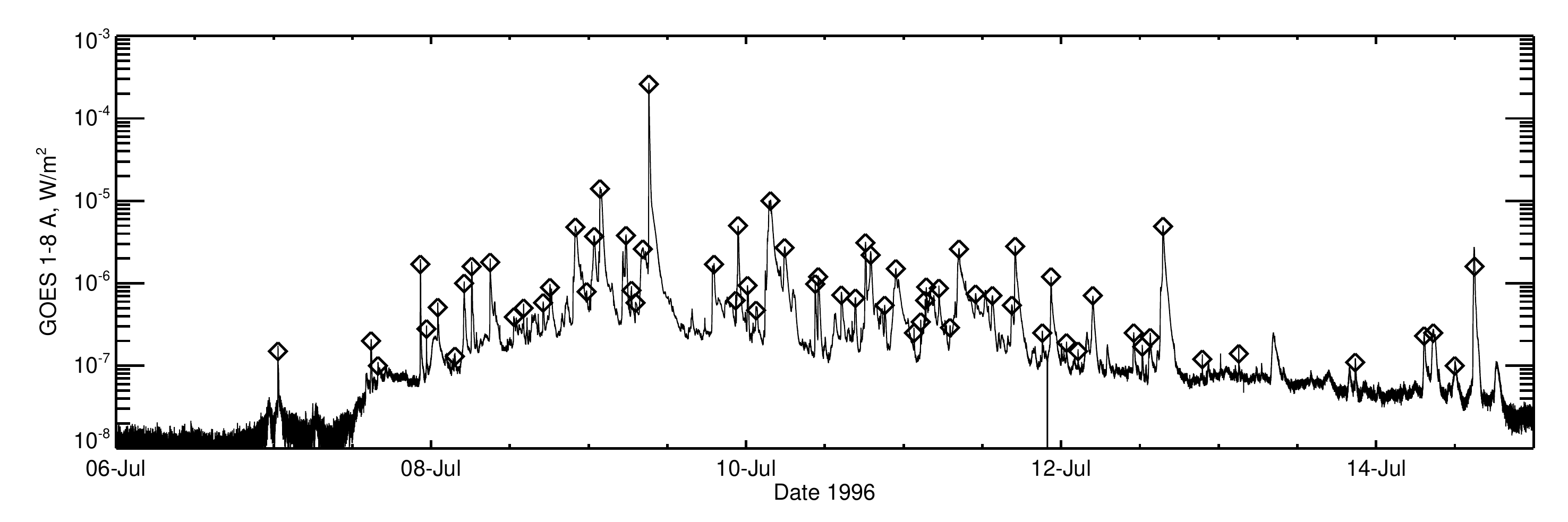}
     \caption{\textit{The GOES soft X-ray time history for NOAA active region~7978, July 1996.
     The diamonds show the listed GOES peak fluxes for flares, all from AR~7978, and a close study reveals some book-keeping errors that inevitably confuse the time-series analysis. Note also the large dynamic range of the soft X-ray flux, which necessitates the usual log scaling here.
     The diamonds show the times and peak fluxes obtained from the NOAA event list, as available in SolarSoft.
      }
} 
\label{fig:ar_07978_goes}
\end{figure*}

\begin{figure*}
\centering
\includegraphics[width=1.5\columnwidth]{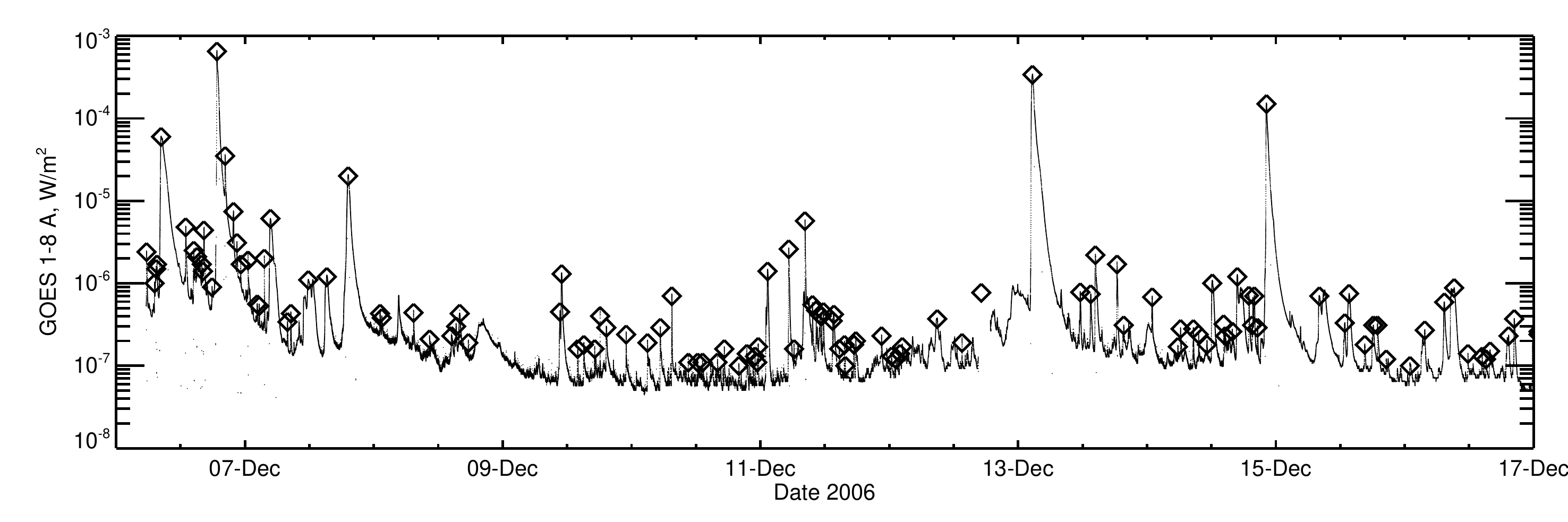}
     \caption{\textit{The GOES soft X-ray time history for NOAA active region~10930, December 2006, in the same format as Figure~\ref{fig:ar_07978_goes}. 
     Again the target region produces all of the listed GOES flare events shown.
     Note that the main flux emergence happened at or prior to the region's E limb passage, in contrast to AR~7978.
      }
} 
\label{fig:ar_01930_goes}
\end{figure*}

Figure~\ref{fig:ar_07978_goes} and Figure~\ref{fig:ar_01930_goes} show the flares used for this study, comparing
the actual GOES time history at full resolution with the database classifications.
The two regions, though both isolated from the point of view of GOES flare production, had different soft X-ray background levels.
The higher background flux for the 2006 region doubtless resulted from the coincidence that its major flux emergence occurred at or before its E~limb arrival, whereas the 1996 region was born near disk center.
The two regions thus sample somewhat different time histories on active-region time scales.
These two figures reveal the approximations involved in the use of a ``data product'' with somewhat unknown properties.
\cite{2005SoPh..227..231W} give a full scientific account of the primary GOES/XRS data. 
In this study we rely solely upon the tabulated NOAA event list, a secondary database generated in near real time, accepting its known and unknown flaws.
The classifications come from an automated detection system with some human intervention, and one can readily identify many omissions and small discrepancies in the two timeseries shown in the Figures.
\nocite{2005SoPh..227..231W}
The database certainly captures every major impulsive event; the GOES spacecraft in principle have 100\% duty cycles.

\section{An Interval-Size Relationship}\label{sec:relationship}

A BUR process should manifest itself as a correlation between the waiting interval and the flare magnitude. 
We can distinguish two cases \citep{2000SoPh..191..381W}: if each flare uses up the entire available stock of free energy, the  interval \textit{before} the flare should correlate with the flare energy; the original \cite{1978ApJ...222.1104R} model predicts this relationship.
Here we term this the ``reset'' limit.
Alternatively, a limited energy release from an existing reservoir could have a correlation between the the time interval \textit{after} the event, on the hypothesis that the same global energy threshold enables the triggering of successive events.
\cite{1998ASSL..229..237H} had suggested this possibility for solar flares, and \cite{2006ApJ...652.1531M} actually discovered such a relationship for the seismic ``glitches'' observed in the period variations of the pulsar PSR J0537--6910, noting that this behavior is not typical for a pulsar.
The glitches arise from sudden adjustments of the neutron-star crust as it evolves on a longer time scale.
For this object the very strong correlation allows observers actually to predict the time of the next glitch.

\begin{figure*}
\centering
\includegraphics[width=\columnwidth]{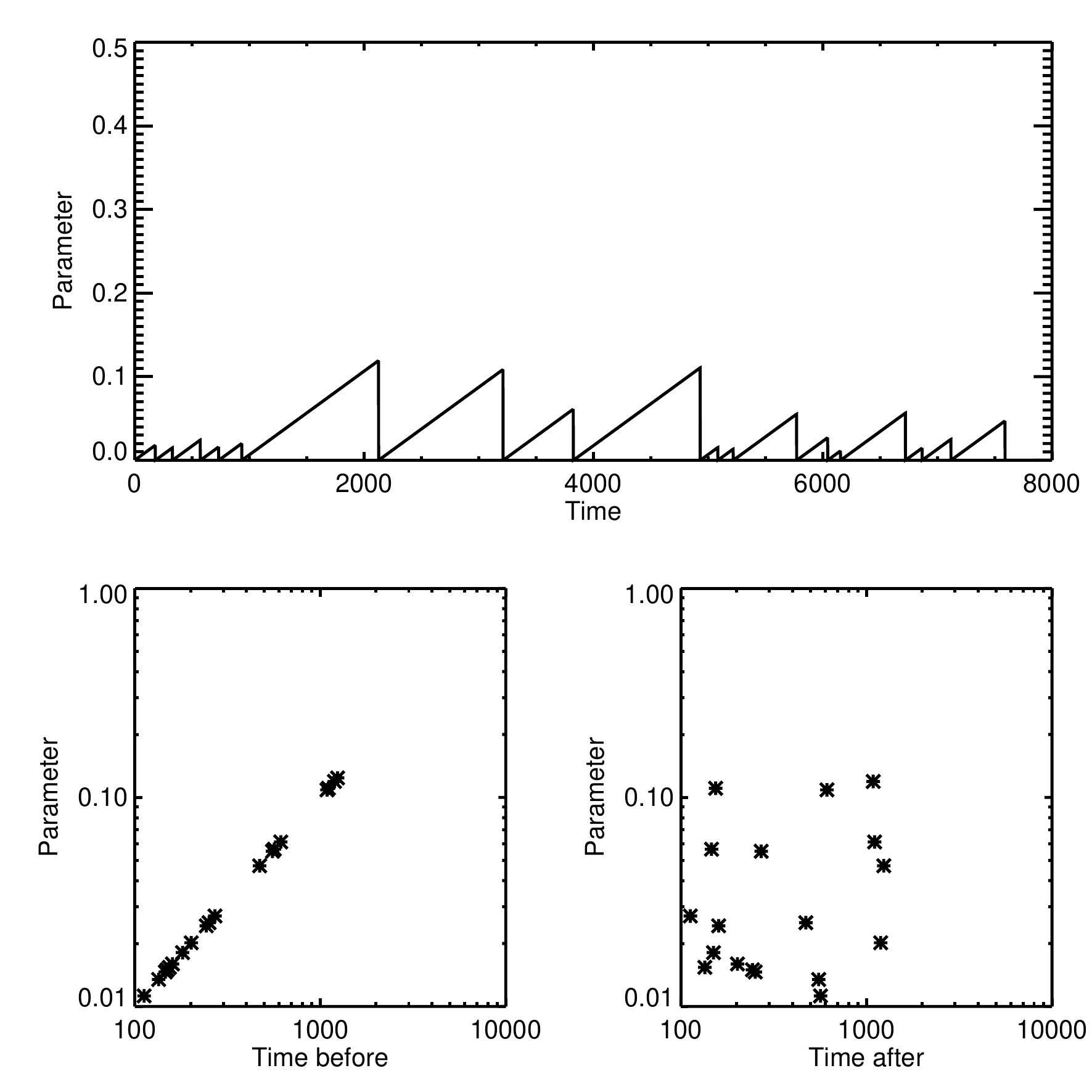}
\includegraphics[width=\columnwidth]{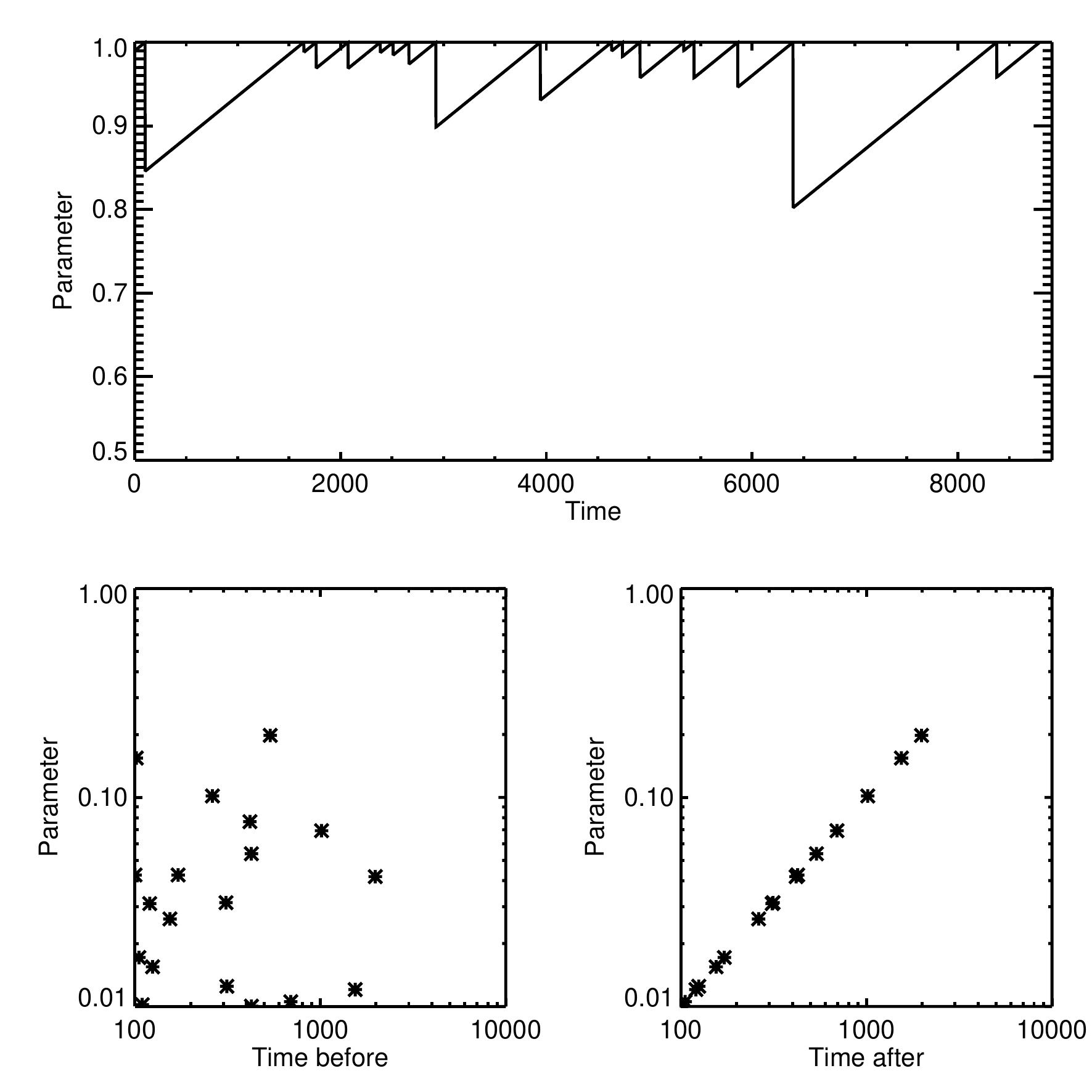}
     \caption{\textit{Toy models for two alternative patterns for an interval-size relationship implying a Build-Up/Release (BUR) scenario \citep{1998ASSL..229..237H}.  
     In the ``reset'' case (left) the parameter builds up gradually to a randomly specified time, and then resets to zero.
     In this case the correlation appears between event magnitude and time \textit{before} the event.
     The ``saturation'' case (right) assumes a fixed non-zero level of the parameter, which acts as a trigger for a subsequent event.
     In this case the correlation appears between event magnitude and time \textit{after} the event.
     The parameter range in the model covers [.01,0.2] in a  power-law distribution with slope $-1.75$, and the occurrence is random within this constraint.
      }
} 
\label{fig:toy}
\end{figure*}

We  construct a toy model with these alternative features, not referring to the basic mathematical description of a relaxation oscillator (the Van der Pol equation); we note that a BUR process might not follow either the ``saturation'' or ``reset'' prescriptions; for example \cite{1995ApJ...447..416L} states that a SOC (Self-Organized Criticality, or ``avalanhe'') model would not have the ``reset'' property (see further discussion in Section~\ref{sec:interp}).
In both cases we consider a steady and constant energy input.
Figure~\ref{fig:toy} illustrates the two alternatives schematically.
Note that although we think of magnetic free energy as the parameter of interest in the model, and use the GOES soft X-ray flux as a proxy for it, this simplification adds variance to any result obtained.
Furthermore, other physical or geometrical parameters of the flaring system may play decisive roles.
We return to discuss this further in Section~\ref{sec:interp} and in the meanwhile, we just describe the toy model  in Figure~\ref{fig:toy} as a generic parameter.

\begin{figure}
\centering
\includegraphics[width=\columnwidth]{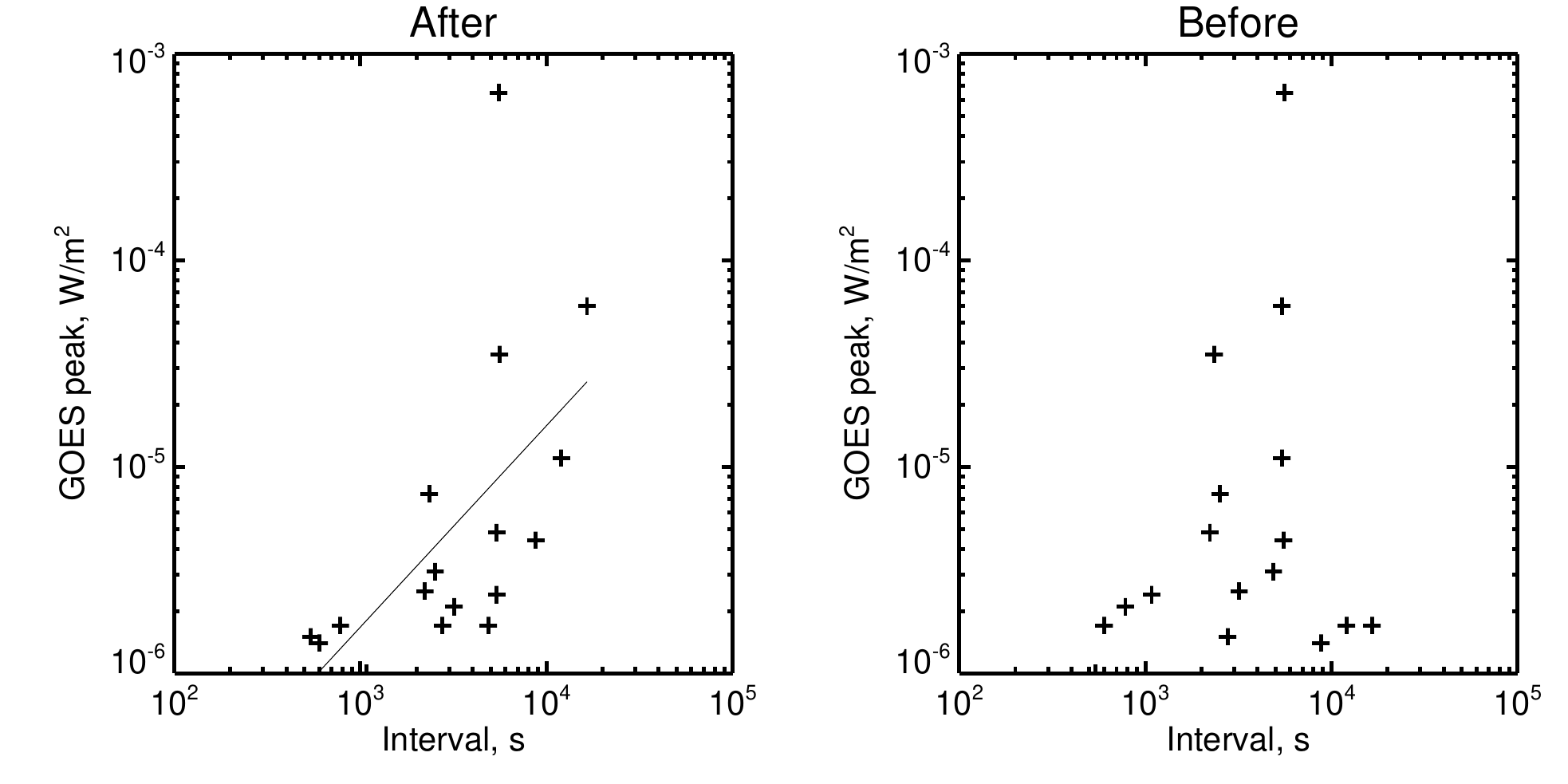}
     \caption{\textit{``saturation'' and ``reset'' correlations for the flare sequence SOL2006-12-06T01 through SOL2006-12-06T23 in NOAA AR~10930.
     The ``saturation'' ordering (left panel) shows a strong correlation, whereas the ``reset'' ordering shows none:
     Pearson correlation coefficients are $0.779 \pm 0.0002$ and $0.13 \pm 0.61$ for ``saturation'' and ``reset'', respectively.
      }
} 
\label{fig:relax_10930_0-16}
\end{figure}

We have searched for ``reset'' and ``saturation'' correlations in the two active regions, starting with the one at the end of Cycle~24, NOAA~AR~10930.
The flare timing reveals a significant ``saturation'' correlation (Figure~\ref{fig:relax_10930_0-16}) during the first day of its flaring life, 6~December 2006, during which an X-class flare occurred. 
Removal of this single point from the correlation did not change the result.
The ``reset'' alternative has no significant correlation.

\begin{figure}
\centering
\includegraphics[width=\columnwidth]{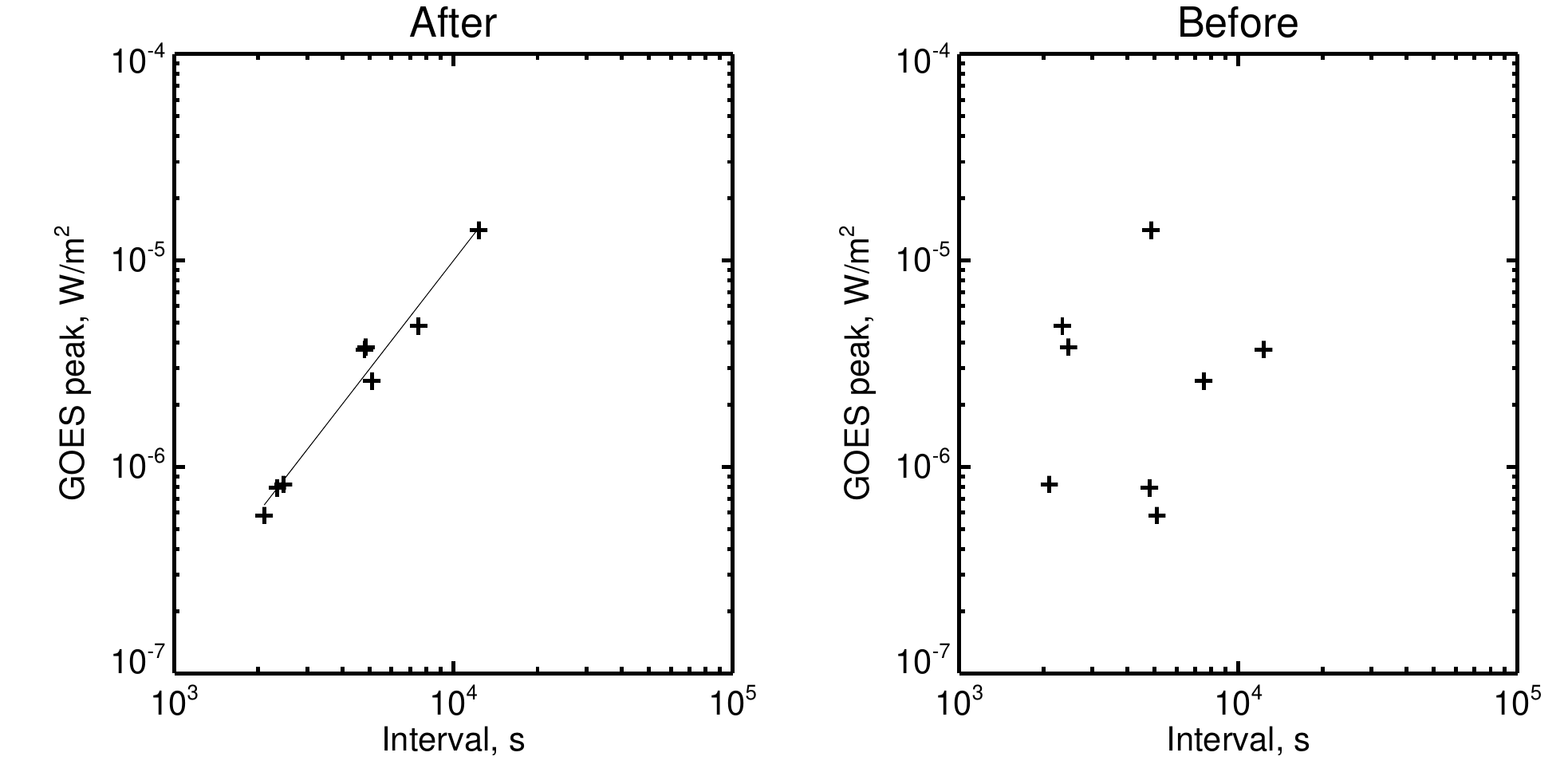}
     \caption{\textit{``Saturation'' and ``reset'' correlations for flares SOL1996-07-08T21 through SOL1996-07-09T09, in NOAA AR~7978 at the end of Cycle 23.
     Again, the ``saturation'' correlation is strong and the ``reset'' correlation non-existent:
        Pearson correlation coefficients are $0.928 \pm 0.001$ and $-0.12 \pm 0.78$ for the ``saturation'' and ``reset'' orderings, respectively.
      }
} 
\label{fig:relax_7978_23-31}
\end{figure}

To confirm this possible correlation, we also re-visited AR~07978, studied inconclusively by \cite{1998ASSL..229..237H}.
Surprisingly, as shown in Figure~\ref{fig:relax_7978_23-31}, one 12-hour interval showed an extremely strong ``saturation'' correlation in this isolated region as well.
Again, the ``reset'' alternative did not correlate, and again the removal of the single X-class flare from the correlation gave the same results.

The correlation results for these two intervals clearly support a relaxation-oscillator (BUR) behavior in the ``saturation'' relationship, and conflict with  the negative conclusions about waiting-time correlations by previous authors \citep{1994PhDT........51B,1998A&A...334..299C,1998ASSL..229..237H,2000ApJ...536L.109W}. 
The difference may results from the systematic uncertainties in the databases in use or in methodology; as discussed below there are many ways to hide the presence of even a strong correlation such as we see in the two examples of Figures~\ref{fig:relax_10930_0-16} and~\ref{fig:relax_7978_23-31}.

\begin{figure*}
\centering
\includegraphics[width=1.9\columnwidth]{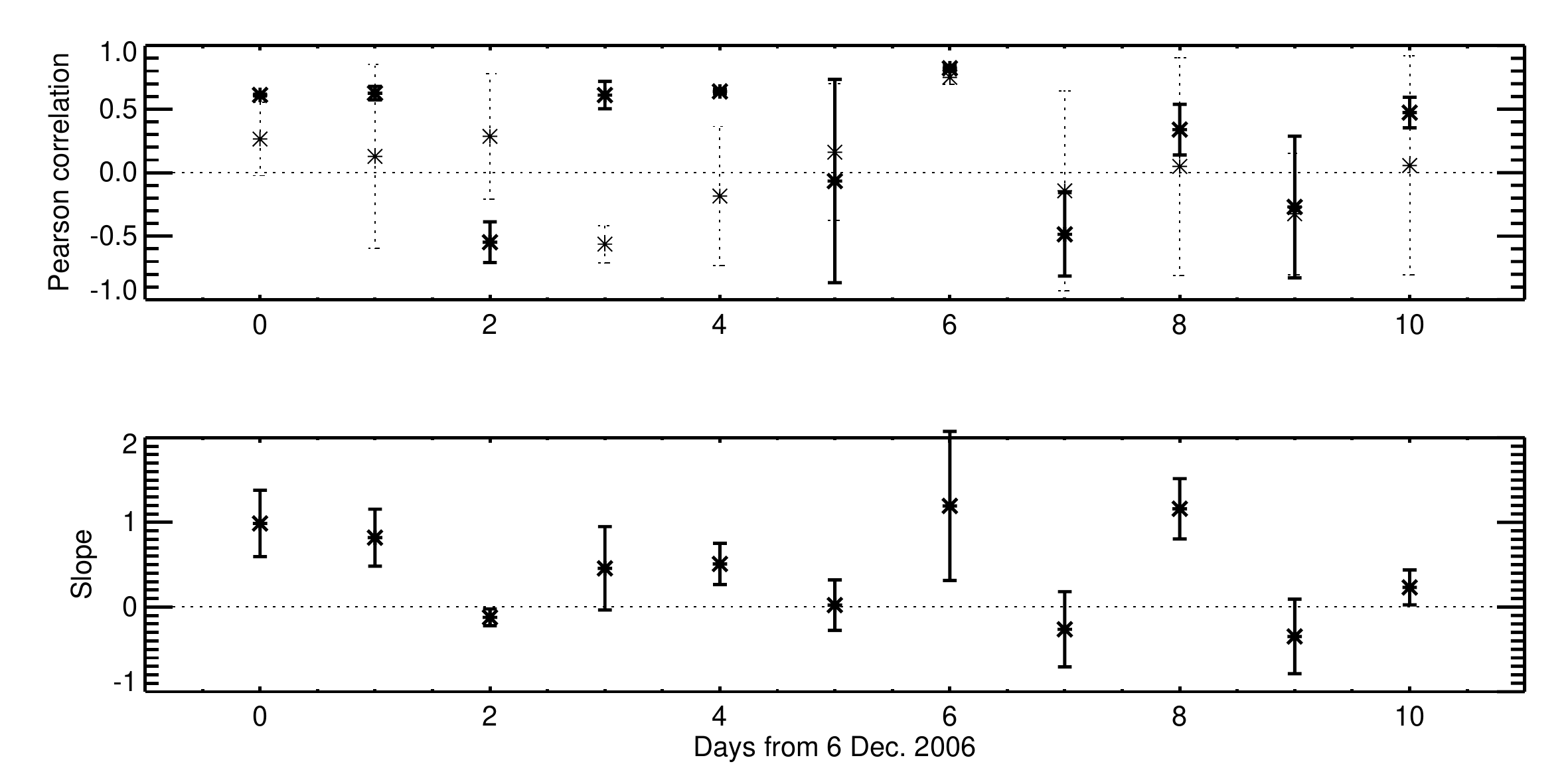}
     \caption{\textit{Upper, Pearson correlation coefficients for both ``saturation'' (solid) and ``reset'' (dotted) correlations for flares on individual days in the disk passage of the isolated region AR 10930.
     In some cases the small uncertainties in the ``saturation' correlations make the ranges too small to see in this graphic.
     Note that while the ``reset'' correlation generally is not significant, the result for 11 December (day 6) is strong for both models.
      }
} 
\label{fig:10930_all}
\end{figure*}

We also use the AR~10930 time series to test the robustness of the ``saturation'' correlation by looking at day-by-day event listings
(Figure~\ref{fig:10930_all}).
Table~\ref{tab:10930_all} also gives the numerical results in the form of Pearson correlation coefficients and their uncertainties ($R_s$ for the ``saturation'' case, and $R_r$ for the ``reset'' case), along with the numbers of flares on each day and the GOES maximum class.
For this single region, the day-by-day analysis shows significant ``saturation'' correlations on several individual days, as corroborated by the power-law fits.
The full time series (123 GOES events from SOL2006-12-06T05:36 through SOL2006-12-17T14:47) showed no correlation, with the Pearson correlation coefficient $r = 0.06 \pm0.48$.
This suggests an intermittent behavior or noise dominance for the correlation on longer time scales, since significant correlations appear on several of the individual days during the disk passage.
In the context of the toy model, a slowly variable driving source would also reduce the correlation in a natural way.
Interestingly, for one day (11 December 2006) we see strong correlations in both senses.

The lower panel of Figure~\ref{fig:10930_all} shows the slope results of linear fits to the day-by-day data from AR~10930, the parameter $\alpha$ in $W \propto (\Delta t)^\alpha$, with $W$ the flare peak flux and $\Delta t$ the waiting time after the flare.
Interestingly the fit uncertainties do not look so convincing as the results for individual days in terms of the Pearson regression coefficients.
Note though that none of the 11 intervals have a significant negative slope.
The slope parameters hint at $\alpha = d(ln S)/d(ln \Delta t) \approx 1$,  but a conclusion about this would 
require a larger sample.

\begin{table}
        \centering
        \caption{Pearson correlation coefficients $R, \delta R$}
        \label{tab:10930_all}
        \begin{tabular}{l r r r l r l}
                \hline
  Date & N & GOES & $R_s$ & $\delta R_s$ & $R_r$ & $\delta R_r$ \\
                \hline
  06-Dec-06 &    18 &  X6.5 &      {\bf 0.61} &      {\bf 0.01}  &      0.26 &      0.29 \\
 07-Dec-06 &    10 &  B4.3 &      {\bf 0.63}  &      {\bf 0.05}  &      0.13 &      0.73 \\
 08-Dec-06 &     8 &  B4.5 &     -0.55 &      0.16 &      0.29 &      0.49 \\
 09-Dec-06 &     8 &  B1.9 &      {\bf 0.61}  &      {\bf 0.11}  &     -0.56 &      0.15 \\
 10-Dec-06 &    13 &  C1.4 &      {\bf 0.64}  &      {\bf 0.02}  &     -0.18 &      0.55 \\
 11-Dec-06 &    17 &  B1.3 &     -0.07 &      0.80 &      0.16 &      0.54 \\
 12-Dec-06 &     7 &  X3.4 &      {\bf 0.82}  &      {\bf 0.02}  &      0.75 &      0.05 \\
 13-Dec-06 &     6 &  B6.8 &     -0.49 &      0.33 &     -0.14 &      0.79 \\
 14-Dec-06 &    16 &  B7.0 &      0.34 &      0.20 &      0.05 &      0.86 \\
 15-Dec-06 &     7 &  B1.0 &     -0.27 &      0.56 &     -0.32 &      0.48 \\
 16-Dec-06 &    12 &  B1.9 &      0.47 &      0.12 &      0.06 &      0.86  \\             
                \hline
        \end{tabular}
\end{table}

\begin{figure*}
\centering
   \includegraphics[width=0.32\textwidth]{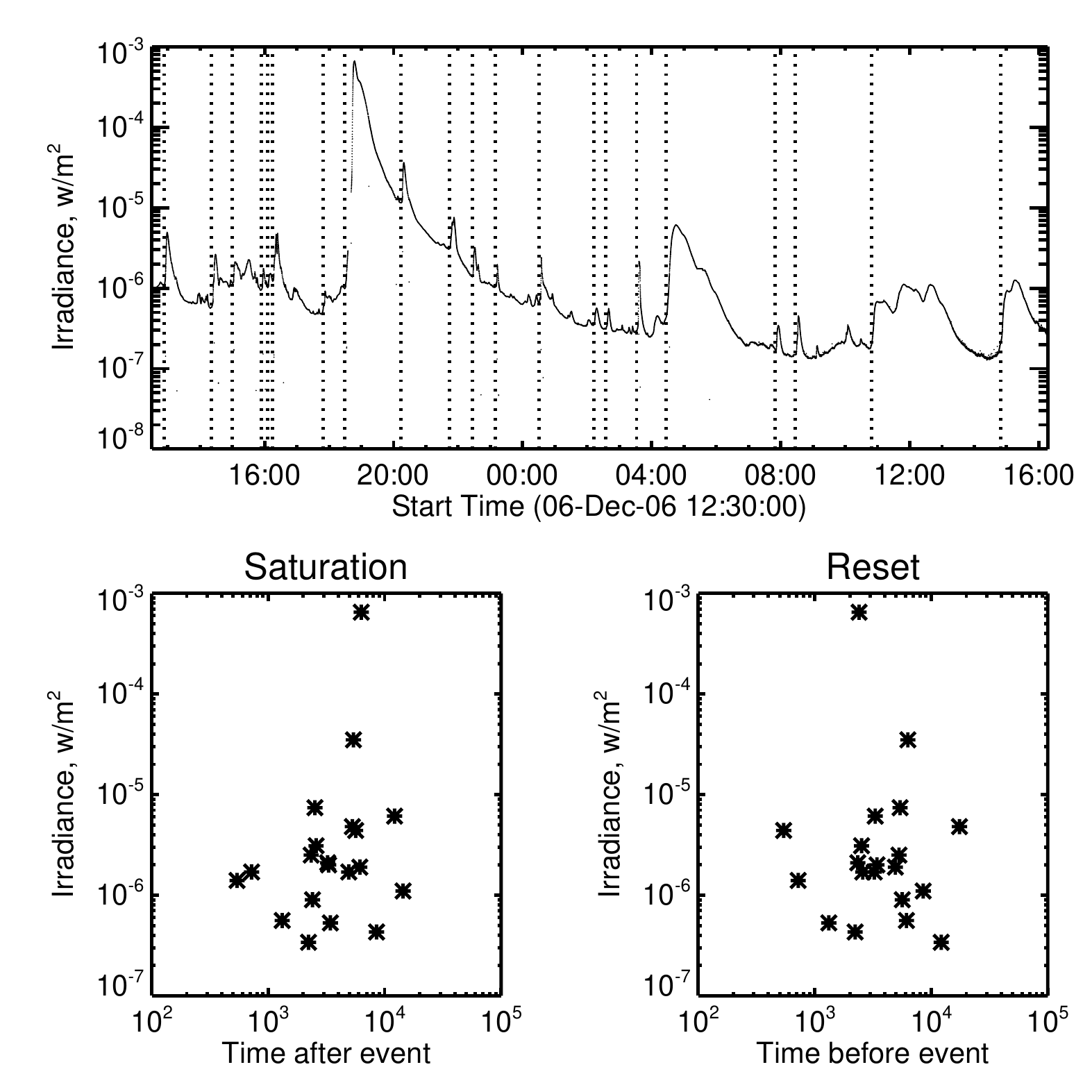}
   \includegraphics[width=0.32\textwidth]{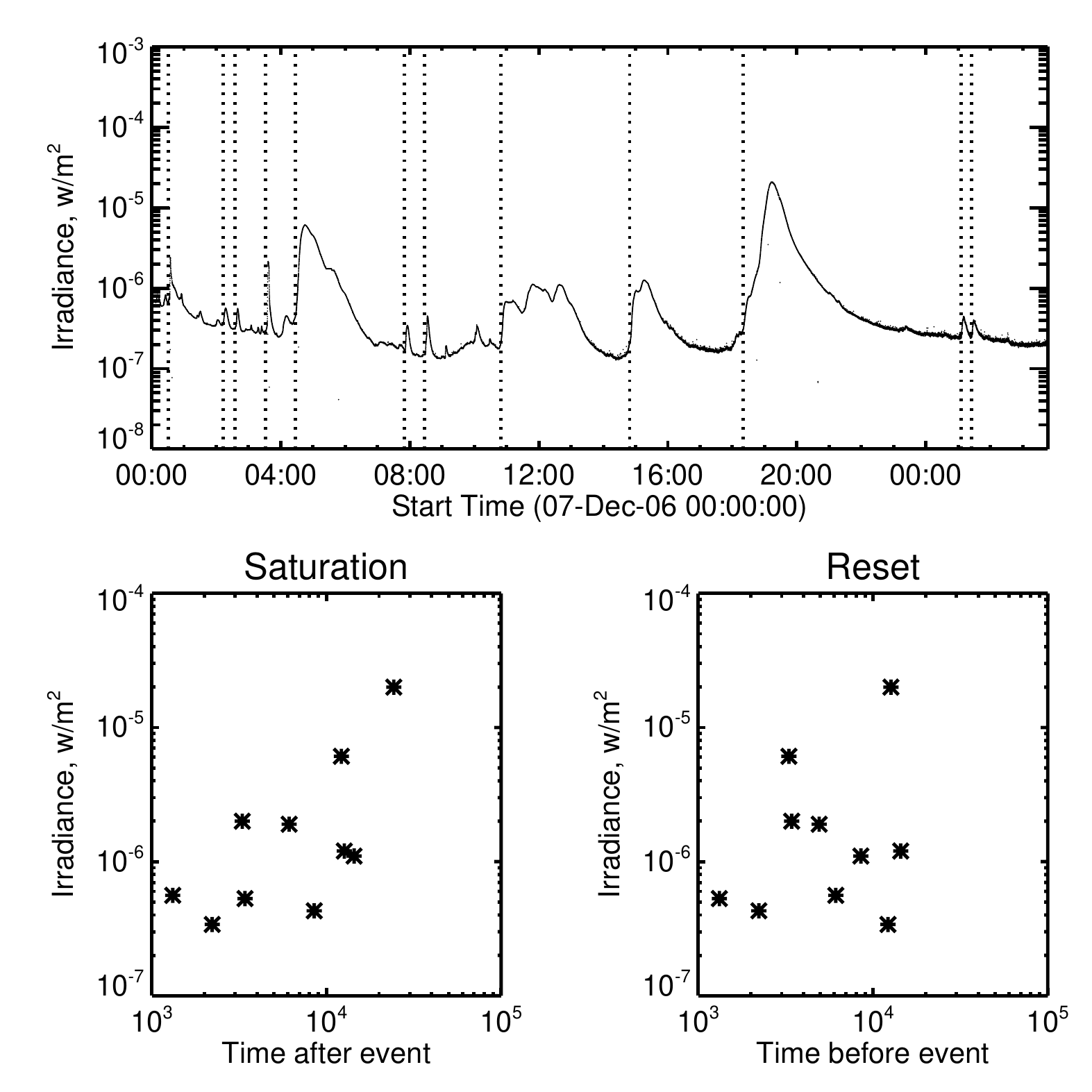}
   \includegraphics[width=0.32\textwidth]{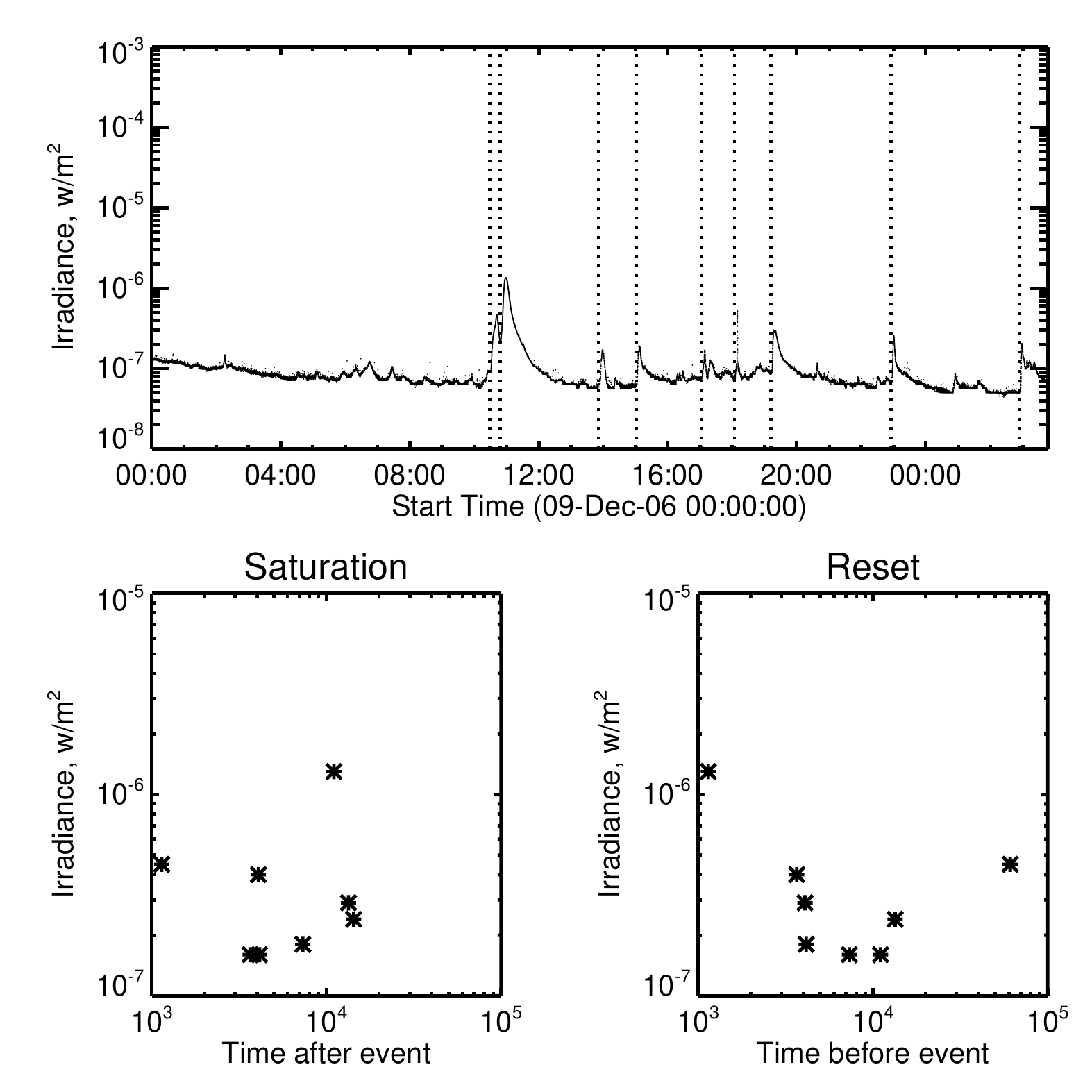}
   \includegraphics[width=0.32\textwidth]{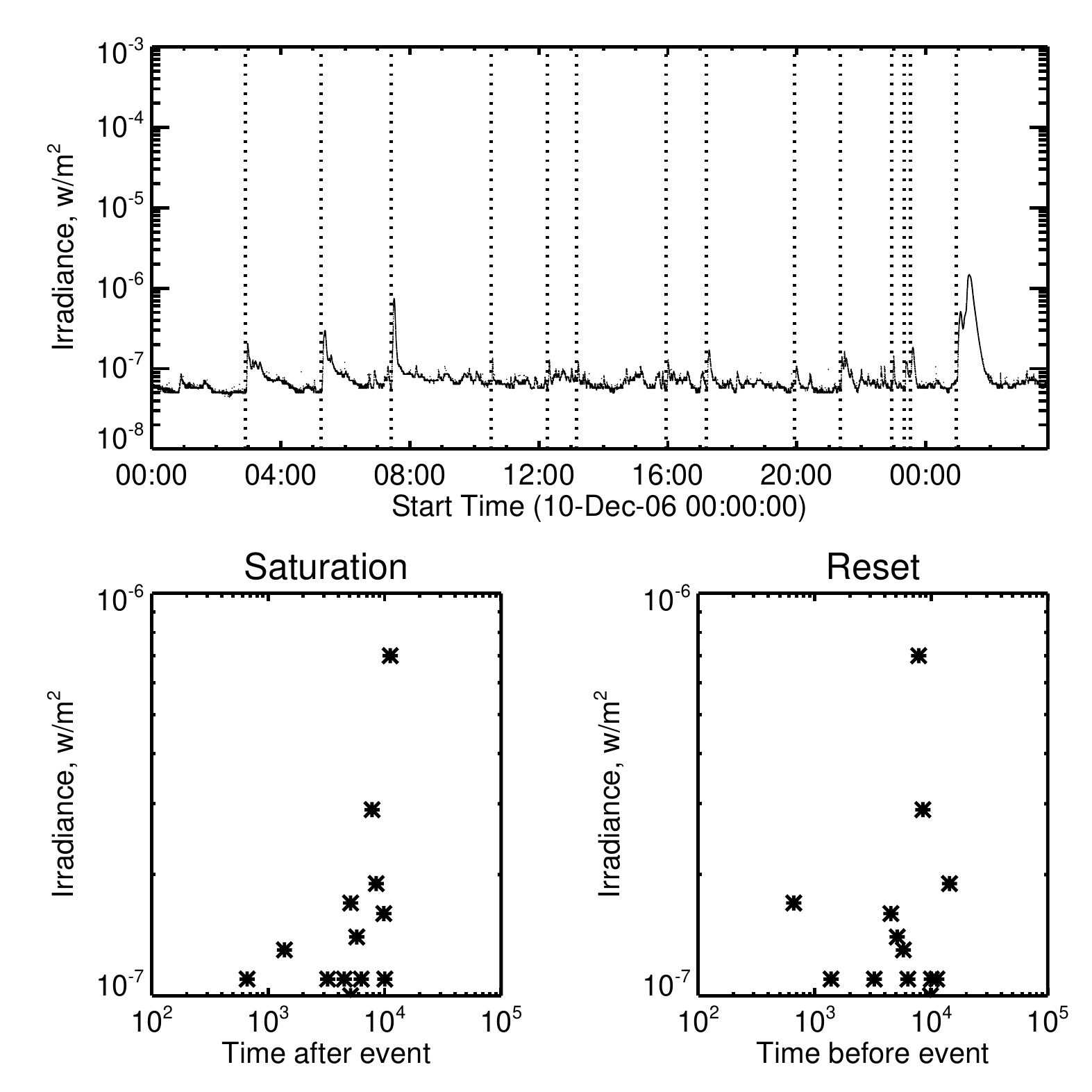}
   \includegraphics[width=0.32\textwidth]{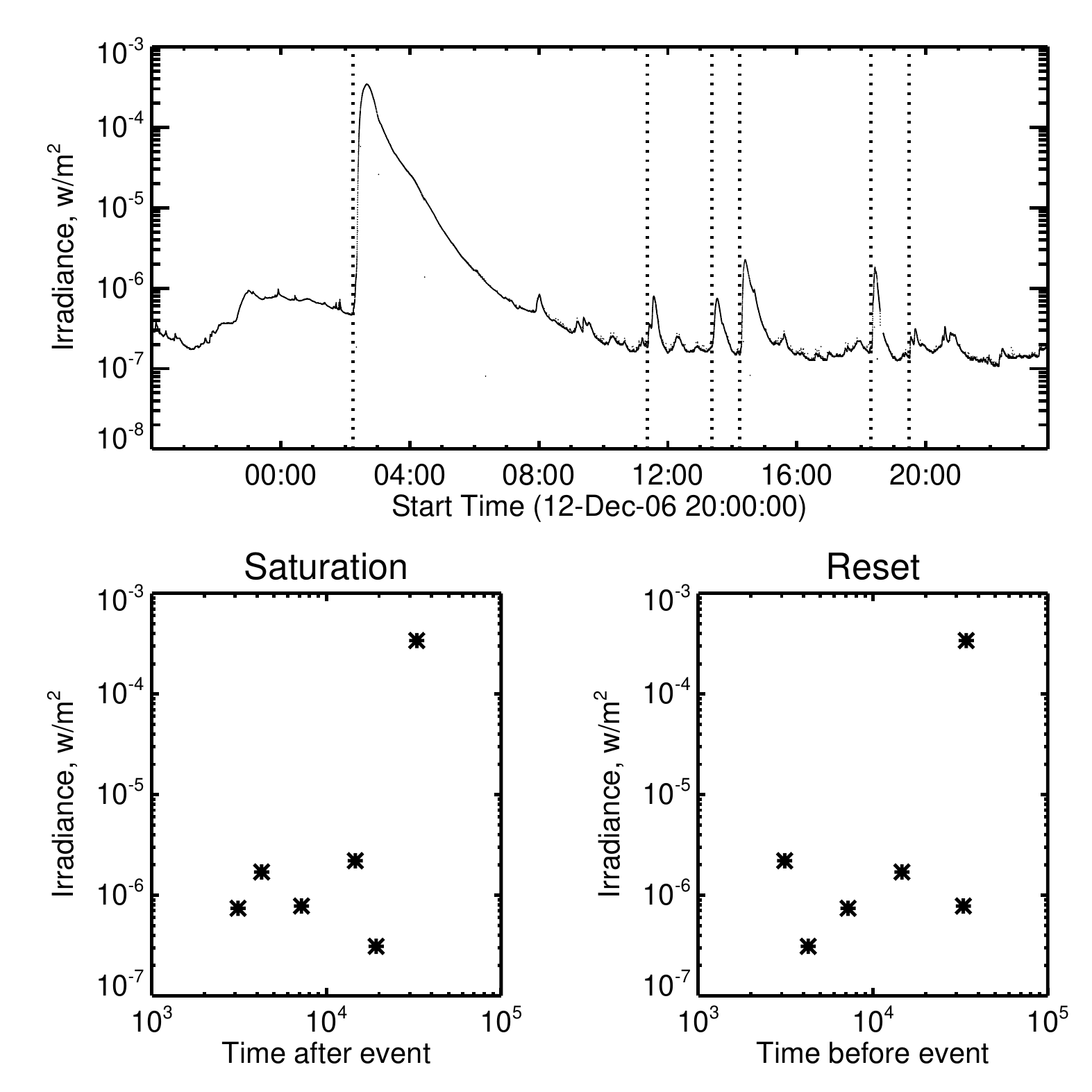}
   \caption{The five individual days with the strongest correlation coefficients in the GOES time history of AR~10930
   (boldface in Table~\ref{tab:10930_all}).
   In each upper panel the vertical dotted lines show the times of the listed events, and for each day the two panels below show the correlation for the ``saturation'' (after) and ``reset'' (before) time intervals.
} 
\label{fig:befaft}
\end{figure*}

Figure~\ref{fig:befaft} shows time series and correlations for the individual one-day intervals with the best correlations for AR~10930 (indicated as boldface in Table~\ref{tab:10930_all}).
Note that these mostly correspond to the one-day intervals with the most energetic events, although one (9~December) peaked at only B1.9, and the previous day had a comparably strong \text{negative} ``saturation'' correlation at and a maximum flare magnitude of B4.5.
The fact that 10~December (maximum C1.4) showed a solid positive ``saturation'' correlation is noteworthy, because a C-class flare should have less obscuration, and we have noted previously that individual correlations remained for intervals with X-class flares even upon removal of their individual entries in the correlation sets.
These results provide strong support for the ``saturation'' correlation.
The bottom line for these views, however, must be that the analysis is close to the limit permitted by random and systematic errors in the use of catalog GOES data; in Section~\ref{sec:concl} we suggest several possible further steps to explore this result.

\section{Why was this correlation not found earlier?}\label{sec:why}

Many sources of uncertainty might mask an interval-size relationship for flare waiting times.
The time reference used here (the GOES soft X-ray peak) has only a crude relationship to the impulsive phase, which might mark the time of the significant energy relase more exactly.
The GOES peak fluxes also do not scale exactly even with the soft X-ray energy, because of variations in flare durations and spectra; another known and probably quite significant scatter also has to come from the routine treatment of the GOES observations here (no background corrections, and no correction made for the ``obscuration'' effect noted for weak events by many \citep[e.g.][]{2001SoPh..203...87W,2014SoPh..289.1341H}.
As noted previously, a long interval may also have reduced correlation in the context of the toy model, if the driving input varies.
In flares with CMEs, a large fraction of the energy \citep[e.g.][]{2012ApJ...759...71E}  may simply disappear from the view of any proxy relating to the flare electromagnetic radiation, and this would produce a substantial bias. 
We could also ask how the heating of the quiescent active region (and Parker's hypothetical nanoflares; see e.g. Cargill \textit{et al.}, 1994) might relate to the coronal reservoir tapped for its flare energy release.
\nocite{1994ApJ...423..854C}

In spite of these systematic errors and unknowns, this study has found a significant correlation by using only the simplest possible flare data, namely the SolarSoft summaries of GOES soft X-ray event time, magnitude, and location.
At least two previous careful searches for interval-size relationships found none, with \cite{1998A&A...334..299C} stating ``No correlation is found between the elapsed time interval between successive flares arising from the same active region and the peak intensity of the flare.''
That study selected sequences of events from the same active region, but perhaps did not explicitly select isolated regions and there may have been no search for the ``saturation'' correlation that we report here, but \cite{2000SoPh..191..381W} used the same database and searched unsuccessfully for both ``reset'' and ``saturation'' matches.
The latter paper offers several possible explanations for the non-detection of the correlation, including the idea that perhaps flare energy is \textit{not} stored visibly in the corona at all!
It also comments interestingly that the existence of a waiting-time correlation would tend to rule out ``avalanche'' models  \citep{1995ApJ...447..416L}.

\section{Interpretation of the ``saturation'' correlation}\label{sec:interp}

The top-right panel of Figure~\ref{fig:toy} suggest the simplest explanation of the ``saturation'' correlation: the available free energy, or some other parameter, builds gradually up to a limit imposed by the active-region structure, at which point a random trigger dislodges a part of the system into a lower-energy state.
The broad distribution of flare magnitudes (generally, a power law or a log-normal observational fit) requires that this end state lie in a continuum; homologous flares sometimes occur but only rarely.
If we identify the generic parameter in the toy model with magnetic free energy in the corona, it presumably consists of inductive storage \citep{1995ApJ...451..391M,1998A&A...337..887Z,2009SSRv..149...83K,2017JGRA..122.7963M} in a system of non-neutralized currents \citep[see e.g.][]{2012ApJ...761...61G}.
Neutralized currents may not store energy so efficiently because of their smaller inductances, but this depends upon the geometry of the system.
The most powerful flare release cannot diminish these currents on short time scales, although it can re-route them and thus alter the inductive energy storage in that way \citep{1995ApJ...451..391M,2012ApJ...748...77S}.
Instead flares appear to leave the active-region structure in a stressed state \citep{1994ApJ...424..436W}, consistent with the ``saturation'' correlation found here.
Thus the total magnetically stored energy sets a firm upper limit on the magnitude of a flare, but the practical (and lower) limit would come from the properties of a ``minimum current corona'' of some kind \citep{1996SoPh..169...91L}.
Either way the power-law distribution of flare energies must roll over around this maximum point, as suggested by observation \citep{1997ApJ...475..338K,2009SoPh..258..141T,2010ApJ...710.1324W} and statistical analysis \citep[e.g.][]{2008SoPh..248...85K}.

\section{Conclusions}\label{sec:concl}

These results presented in this paper match expectations for a build-up/release scenario for solar flares occurring in isolated active regions, which if confirmed would establish a pattern widely believed in, but not previously established observationally.
The result found suggests a ``saturation'' ordering, with correlated time intervals \textit{after} the event, rather than the ``reset'' relaxation often discussed, as illustrated in Figure~\ref{fig:toy}.
The correlation may persist for a time scale on the order of a day, and it exhibits intermittency, some of which must result from random and systematic error over epochs with limited ranges of flare magnitude.
The triggering of the events remains as a ``piecewise Poisson'' process \citep{2000ApJ...536L.109W}.
The unique parameter dictating the ``saturation'' level appears could be the available free magnetic energy, since our search has been in terms of a known correlate, the GOES soft X-ray peak flux.
Physically we expect that helicity may also play a key role \citep[e.g.,][]{1994GeoRL..21..241R}, but our descriptive toy model only needs a single parameter. 

The appearance of the ``saturation'' correlation in a crude analysis (we have only used imprecise catalog information) suggests several further lines of research to check and possibly extend the relationship:
\begin{itemize}
\item A more thorough search of the GOES statistics, based on a quantitative analysis considering uncertainties; this would require using the primary data, rather than the existing catalog.
\item Searches with hard X-rays, which have much smaller obscuration in the time domain; here RHESSI data could be invaluable because of its image capability (note that its low Earth orbit restricts the uninterrupted time range).
\item Time-interval analysis of microflare activity within a given active region, as observed with imaging instruments \citep[e.g.][]{1995PASJ...47..251S}.
\item Comparisons with image-based analysis of Poynting flux and helicity transport in specific active regions, specifically in well-documented homologous flare sequences \citep[e.g.,][]{2019SoPh..294....4R}.
\item Theoreticawork on feasible model descriptions, attempting to understand how the apparent one-parameter instability limit can lead to a continuum of final states.
Avalanche models \citep{1991ApJ...380L..89L} remain interesting \citep[e.g.,][]{2018A&A...615A..84R,2019arXiv190900195F} even if they have not thus far anticipated the result suggested here.
\end{itemize}

More thorough studies of flare waiting times may resolve several interesting questions involving CME occurrence, interactions between regions, time scales associated with trans-photospheric Poynting flux, \textit{etc}.
The results may also offer the possibility of improving short-term prediction of flare magnitudes based on the interpretation of prior flare occurrence to reflect the magnitude of the Poynting flux responsible for the energy build-up in a specific region, noting the empirical success of the ``after'' correlation in anticipating pulsar glitches in PSR J0537--6910 \citep{2006ApJ...652.1531M,2018ApJ...863..196M}.

\bigskip
\noindent{\bf Acknowledgment:} A remark by Manolis Georgoulis, noting the unambiguous flare time series from the isolated active region NOAA 10930, provided the impetus for this study, which follows on from preliminary work I had done in 1997 \citep{1998ASSL..229..237H}; please see \url{http://www.ssl.berkeley.edu/~hhudson/publications/1998ASSL..229..237H.pdf} for that paper, which may be difficult to obtain otherwise. I would also like to thank Doug Biesecker, Kris Cooper, Lyndsay Fletcher, Manolis Georgoulis, Brian Welsch, Mike Wheatland, and Graham Woan for helpful discussions.

\bibliography{relax}
\bibliographystyle{mnras}

\end{document}